\documentstyle[12pt]{article}
\skewchar\fivmi='177
\skewchar\sixmi='177
\skewchar\sevmi='177
\skewchar\egtmi='177
\skewchar\ninmi='177
\skewchar\tenmi='177
\skewchar\elvmi='177
\skewchar\twlmi='177
\skewchar\frtnmi='177
\skewchar\svtnmi='177
\skewchar\twtymi='177
\def\@magscale#1{ scaled \magstep #1}
\font\twfvmi  = ammi10   \@magscale5 
\skewchar\twfvmi='177
\skewchar\fivsy='60
\skewchar\sixsy='60
\skewchar\sevsy='60
\skewchar\egtsy='60
\skewchar\ninsy='60
\skewchar\tensy='60
\skewchar\elvsy='60
\skewchar\twlsy='60
\skewchar\frtnsy='60
\skewchar\svtnsy='60
\skewchar\twtysy='60
\font\twfvsy  = amsy10   \@magscale5 
\skewchar\twfvsy='60

\catcode`@=11
\def\un#1{\relax\ifmmode\@@underline#1\else
        $\@@underline{\hbox{#1}}$\relax\fi}
\catcode`@=12

\let\du=\d                      
\let\um=\H                      

\def\a{\alpha}
\def\b{\beta}

\def\d{\delta}
\def\e{\epsilon}

\def\l{\lambda}

\def\q{\theta}
\def\r{\rho}

\def\G{\Gamma}

\font\sc=font005                        
\def\Sc#1{{\hbox{\sc #1}}}      
\font\ooo=circle10                      
\font\ro=manfnt                         
\def\kcl{{\hbox{\ro 6}}}                
\def\kcr{{\hbox{\ro 7}}}                
\def\ktl{{\hbox{\ro \char'134}}}        
\def\ktr{{\hbox{\ro \char'135}}}        
\def\kbl{{\hbox{\ro \char'136}}}        
\def\kbr{{\hbox{\ro \char'137}}}        

\def\ip{{=\!\!\! \mid}}                                    
\def\bo{{\raise.15ex\hbox{\large$\Box$}}}               
\def\pr{\prod}                                          
\def\TH{{\raise.2ex\hbox{$\displaystyle \bigodot$}\mskip-4.7mu \llap H \;}}
\def\face{{\raise.2ex\hbox{$\displaystyle \bigodot$}\mskip-2.2mu \llap {$\ddot
        \smile$}}}                                      

\def\sp#1{{}^{#1}}                              
\def\Tilde#1{{\widetilde{#1}}\hskip 0.03in}                     
\def\Hat#1{\widehat{#1}}                        
\def\Bar#1{\overline{#1}}                       
\def\leftrightarrowfill{$\mathsurround=0pt \mathord\leftarrow \mkern-6mu
        \cleaders\hbox{$\mkern-2mu \mathord- \mkern-2mu$}\hfill
        \mkern-6mu \mathord\rightarrow$}
\def\dvec#1{\vbox{\ialign{##\crcr
        \leftrightarrowfill\crcr\noalign{\kern-1pt\nointerlineskip}
        $\hfil\displaystyle{#1}\hfil$\crcr}}}           
\def\dt#1{{\buildrel {\hbox{\LARGE .}} \over {#1}}}     

\def\frac#1#2{{\textstyle{#1\over\vphantom2\smash{\raise.20ex
        \hbox{$\scriptstyle{#2}$}}}}}                   
\def\ha{\frac12}                                        
\def\sfrac#1#2{{\vphantom1\smash{\lower.5ex\hbox{\small$#1$}}\over
        \vphantom1\smash{\raise.4ex\hbox{\small$#2$}}}} 
\def\bfrac#1#2{{\vphantom1\smash{\lower.5ex\hbox{$#1$}}\over
        \vphantom1\smash{\raise.3ex\hbox{$#2$}}}}       
\def\afrac#1#2{{\vphantom1\smash{\lower.5ex\hbox{$#1$}}\over#2}}    

\newskip\humongous \humongous=0pt plus 1000pt minus 1000pt
\def\caja{\mathsurround=0pt}
\def\eqalign#1{\,\vcenter{\openup2\jot \caja
        \ialign{\strut \hfil$\displaystyle{##}$&$
        \displaystyle{{}##}$\hfil\crcr#1\crcr}}\,}
\newif\ifdtup
\def\panorama{\global\dtuptrue \openup2\jot \caja
        \everycr{\noalign{\ifdtup \global\dtupfalse
        \vskip-\lineskiplimit \vskip\normallineskiplimit
        \else \penalty\interdisplaylinepenalty \fi}}}
\def\li#1{\panorama \tabskip=\humongous                         
        \halign to\displaywidth{\hfil$\displaystyle{##}$
        \tabskip=0pt&$\displaystyle{{}##}$\hfil
        \tabskip=\humongous&\llap{$##$}\tabskip=0pt
        \crcr#1\crcr}}

\def\ref#1{$\sp{#1)}$}

\parskip=\medskipamount                 
\thicklines                         

\def\oldheadpic{                                
        \setlength{\unitlength}{.4mm}
        \thinlines
        \par
        \begin{picture}(349,16)
        \put(325,16){\line(1,0){4}}
        \put(330,16){\line(1,0){4}}
        \put(340,16){\line(1,0){4}}
        \put(335,0){\line(1,0){4}}
        \put(340,0){\line(1,0){4}}
        \put(345,0){\line(1,0){4}}
        \put(329,0){\line(0,1){16}}
        \put(330,0){\line(0,1){16}}
        \put(339,0){\line(0,1){16}}
        \put(340,0){\line(0,1){16}}
        \put(344,0){\line(0,1){16}}
        \put(345,0){\line(0,1){16}}
        \put(329,16){\oval(8,32)[bl]}
        \put(330,16){\oval(8,32)[br]}
        \put(339,0){\oval(8,32)[tl]}
        \put(345,0){\oval(8,32)[tr]}
        \end{picture}
        \par
        \thicklines
        \vskip.2in}
\def\oldtitle#1#2#3#4{\oldheadpic\begin{center}\vglue.5in{\large\bf #1}\\[.6in]
        {#2}\\[.1in] {\it Department of Physics and Astronomy}\\
        {\it University of Maryland, College Park, MD 20742}\\[.6in]
        Physics Publication \#{#3}\\ {#4}\\[1.5in] {\bf Abstract}\\[.1in]
        \end{center} \begin{quotation}}                 
\def\oldTitle#1#2#3#4#5#6#7{\oldheadpic\begin{center} \vglue .4in
        {\large\bf #1}\\[.4in]
        {#2}\\[.1in] {\it Department of Physics and Astronomy}\\
        {\it University of Maryland, College Park, MD 20742}\\[.1in]
        {#3}\\[.1in] {\it {#4}}\\ {\it {#5}}\\[.4in]
        Physics Publication \#{#6}\\ {#7}\\[.5in] {\bf Abstract}\\[.1in]
        \end{center} \begin{quotation}}                 
\def\border{                                            
        \setlength{\unitlength}{1mm}
        \newcount\xco
        \newcount\yco
        \xco=-24
        \yco=12
        \begin{picture}(140,0)
        \put(\xco,\yco){$\ktl$}
        \advance\yco by-1
        {\loop
        \put(\xco,\yco){$\kcl$}
        \advance\yco by-2
        \ifnum\yco>-240
        \repeat
        \put(\xco,\yco){$\kbl$}}
        \xco=158
        \yco=12
        \put(\xco,\yco){$\ktr$}
        \advance\yco by-1
        {\loop
        \put(\xco,\yco){$\kcr$}
        \advance\yco by-2
        \ifnum\yco>-240
        \repeat
        \put(\xco,\yco){$\kbr$}}
        \put(-20,11){\tiny University of Maryland Elementary Particle
Physics University of Maryland Elementary Particle Physics University of
Maryland Elementary Particle Physics}
        \put(-20,-241.5){\tiny University of Maryland Elementary
Particle Physics University of Maryland Elementary Particle Physics
University of Maryland Elementary Particle Physics}
        \end{picture}
        \par\vskip-8mm}
\def\bordero{                                           
        \setlength{\unitlength}{1mm}
        \newcount\xco
        \newcount\yco
        \xco=-24
        \yco=12
        \begin{picture}(140,0)
        \put(\xco,\yco){$\ktl$}
        \advance\yco by-1
        {\loop
        \put(\xco,\yco){$\kcl$}
        \advance\yco by-2
        \ifnum\yco>-240
        \repeat
        \put(\xco,\yco){$\kbl$}}
        \xco=158
        \yco=12
        \put(\xco,\yco){$\ktr$}
        \advance\yco by-1
        {\loop
        \put(\xco,\yco){$\kcr$}
        \advance\yco by-2
        \ifnum\yco>-240
        \repeat
        \put(\xco,\yco){$\kbr$}}
        \put(-20,12){\ooo
bacdefghidfghghdhededbihdgdfdfhhdheidhdhebaaahjhhdahba
hgdedge
   hgfdiehhgdigicba}
        \put(-20,-241.5){\ooo
ababaighefdbfghgeahgdfgafagihdidihiidhiagfedhadbfd
ecdcdfa
   gdcbhaddhbgfchbgfdacfediacbabab}
        \end{picture}
        \par\vskip-8mm}
\def\headpic{                                           
        \indent
        \setlength{\unitlength}{.4mm}
        \thinlines
        \par
        \begin{picture}(29,16)
        \put(165,16){\line(1,0){4}}
        \put(170,16){\line(1,0){4}}
        \put(180,16){\line(1,0){4}}
        \put(175,0){\line(1,0){4}}
        \put(180,0){\line(1,0){4}}
        \put(185,0){\line(1,0){4}}
        \put(169,0){\line(0,1){16}}
        \put(170,0){\line(0,1){16}}
        \put(179,0){\line(0,1){16}}
        \put(180,0){\line(0,1){16}}
        \put(184,0){\line(0,1){16}}
        \put(185,0){\line(0,1){16}}
        \put(169,16){\oval(8,32)[bl]}
        \put(170,16){\oval(8,32)[br]}
        \put(179,0){\oval(8,32)[tl]}
        \put(185,0){\oval(8,32)[tr]}
        \end{picture}
        \par\vskip-6.5mm
        \thicklines}
\def\title#1#2#3#4{\border\headpic {\hbox to\hsize{#4 \hfill UMDEPP #3}}\par
        \begin{center} \vglue .5in {\large\bf #1}\\[.6in]
        {#2}\\[.1in] {\it Department of Physics and Astronomy}\\
        {\it University of Maryland, College Park, MD 20742}\\[1.5in]
        {\bf Abstract}\\[.1in] \end{center} \begin{quotation}}  
\def\Title#1#2#3#4#5#6#7{\border\headpic
        {\hbox to\hsize{#7 \hfill UMDEPP #6}}\par
        \begin{center} \vglue .4in {\large\bf #1}\\[.4in]
        {#2}\\[.1in] {\it Department of Physics and Astronomy}\\
        {\it University of Maryland, College Park, MD 20742}\\[.1in]
        {#3}\\[.1in] {\it {#4}}\\ {\it {#5}}\\[.5in] {\bf Abstract}\\[.1in]
        \end{center} \begin{quotation}}                 
\def\endtitle{\end{quotation}\newpage}                  

\def\sect#1{\bigskip\medskip \goodbreak \noindent{\bf {#1}} \nobreak \medskip}
\def\refs{\sect{References} \footnotesize \frenchspacing \parskip=0pt}
\def\Item{\par\hang\textindent}

\begin{document}

\def\gg{{\hbox{\sc g}}}
\def\nt{$~N=2$~}
\def\gg{{\hbox{\sc g}}}
\def\nt{$~N=2$~}
\def\tr{{\rm tr}}
\def\Tr{{\rm Tr}}
\def\mpl#1#2#3{Mod.~Phys.~Lett.~{\bf A{#1}} (19{#2}) #3}

\def\.{.$\,$}
\def\scst{\scriptstyle}
\def\itrema{$\ddot{\scriptstyle 1}$}
\def\Bo{\bo{\hskip 0.03in}}
\def\lrad#1{ \left( A {\buildrel\leftrightarrow\over D}_{#1} B\right) }
\def\derx{\partial_x} \def\dery{\partial_y} \def\dert{\partial_t}
\def\Vec#1{{\overrightarrow{#1}}}

\def\ula{{\underline a}} \def\ulb{{\underline b}} \def\ulc{{\underline c}}
\def\uld{{\underline d}} \def\ule{{\underline e}} \def\ulf{{\underline f}}
\def\ulg{{\underline g}} \def\ulm{{\underline m}}
\def\uln#1{\underline{#1}}
\def\ulp{{\underline p}} \def\ulq{{\underline q}} \def\ulr{{\underline r}}

\def\plpl{{+\!\!\!\!\!{\hskip 0.009in}{\raise -1.0pt\hbox{$_+$}}
{\hskip 0.0008in}}}

\def\mimi{{-\!\!\!\!\!{\hskip 0.009in}{\raise -1.0pt\hbox{$_-$}}
{\hskip 0.0008in}}}

\def\items#1{\\ \item{[#1]}}
\def\ul{\underline}
\def\un{\underline}
\def\-{{\hskip 1.5pt}\hbox{-}}

\def\kd#1#2{\d\du{#1}{#2}}
\def\fracmm#1#2{{{#1}\over{#2}}}
\def\footnotew#1{\footnote{\hsize=6.5in {#1}}}

\def\low#1{{\raise -3pt\hbox{${\hskip 1.0pt}\!_{#1}$}}}

\def\ip{{=\!\!\! \mid}}
\def\unb{{\underline {\bar n}}}
\def\upb{{\underline {\bar p}}}
\def\um{{\underline m}}
\def\up{{\underline p}}
\def\Phib{{\Bar \Phi}}
\def\Phit{{\tilde \Phi}}
\def\Phibt{{\tilde {\Bar \Phi}}}
\def\Db{{\Bar D}_{+}}
\def\gg{{\hbox{\sc g}}}
\def\nt{$~N=2$~}

\border\headpic {\hbox to\hsize{March, 1993\hfill UMDEPP 93--145}}\par
\begin{center}
\vglue .25in

{\large\bf Supersymmetric KP Systems Embedded in} \\
{\large\bf Supersymmetric Self--Dual Yang--Mills Theory}$\,$\footnote{This
work is supported in part by NSF grant \# PHY-91-19746.} \\[.1in]

\baselineskip 10pt

\vskip 0.25in

Hitoshi NISHINO\footnote{E-Mail: Nishino@UMDHEP.bitnet} \\[.2in]
{\it Department of Physics} \\ [.015in]
{\it University of Maryland at College Park}\\ [.015in]
{\it College Park, MD 20742-4111, USA} \\[.1in]
and\\[.1in]
{\it Department of Physics and Astronomy} \\[.015in]
{\it Howard University} \\[.015in]
{\it Washington, D.C. 20059, USA} \\[.18in]

\vskip 1.0in

{\bf Abstract}\\[.1in]
\end{center}

\begin{quotation}

{}~~~We show that $~N=1$~ {\it supersymmetric} Kadomtsev-Petviashvili
(SKP) equations can be embedded into recently formulated $~N=1$~
self-dual {\it supersymmetric} Yang-Mills theories after appropriate
dimensional reduction and truncation, which yield three-dimensional
supersymmetric Chern-Simons
theories.  Based on this result, we also give conjectural
\hbox{$N=2~$} SKP equations.  Subsequently some exact solutions of these
systems including fermionic fields are given.

\endtitle

\def\doit#1#2{\ifcase#1\or#2\fi}
\def\[{\lfloor{\hskip 0.35pt}\!\!\!\lceil}
\def\]{\rfloor{\hskip 0.35pt}\!\!\!\rceil}
\def\delsl{{{\partial\!\!\! /}}}
\def\caldsl{{\calD\!\!\! /}}
\def\calO{{\cal O}}
\def\asym{({\scriptstyle 1\leftrightarrow \scriptstyle 2})}
\def\Lag{{\cal L}}
\def\du#1#2{_{#1}{}^{#2}}
\def\ud#1#2{^{#1}{}_{#2}}
\def\dud#1#2#3{_{#1}{}^{#2}{}_{#3}}
\def\udu#1#2#3{^{#1}{}_{#2}{}^{#3}}
\def\calD{{\cal D}}
\def\calM{{\cal M}}
\def\tildef{{\tilde f}}
\def\calDsl{{\calD\!\!\!\! /}}

\def\Hat#1{{#1}{\large\raise-0.02pt\hbox{$\!\hskip0.038in\!\!\!\hat{~}$}}}
\def\hati{{\hat{I}}}
\def\dt{$~D=10$~}
\def\alp{\alpha{\hskip 0.007in}'}
\def\oalp#1{\alp^{\hskip 0.007in {#1}}}
\def\naive{{{na${\scriptstyle 1}\!{\dot{}}\!{\dot{}}\,\,$ve}}}
\def\items#1{\vskip 0.05in\Item{[{#1}]}}
\def\item#1{\Item{#1}}

\def\pl#1#2#3{Phys.~Lett.~{\bf {#1}B} (19{#2}) #3}
\def\np#1#2#3{Nucl.~Phys.~{\bf B{#1}} (19{#2}) #3}
\def\prl#1#2#3{Phys.~Rev.~Lett.~{\bf #1} (19{#2}) #3}
\def\pr#1#2#3{Phys.~Rev.~{\bf D{#1}} (19{#2}) #3}
\def\cqg#1#2#3{Class.~and Quant.~Gr.~{\bf {#1}} (19{#2}) #3}
\def\cmp#1#2#3{Comm.~Math.~Phys.~{\bf {#1}} (19{#2}) #3}
\def\jmp#1#2#3{Jour.~Math.~Phys.~{\bf {#1}} (19{#2}) #3}
\def\ap#1#2#3{Ann.~of Phys.~{\bf {#1}} (19{#2}) #3}
\def\prep#1#2#3{Phys.~Rep.~{\bf {#1}C} (19{#2}) #3}
\def\ptp#1#2#3{Prog.~Theor.~Phys.~{\bf {#1}} (19{#2}) #3}
\def\ijmp#1#2#3{Int.~Jour.~Mod.~Phys.~{\bf {#1}} (19{#2}) #3}
\def\nc#1#2#3{Nuovo Cim.~{\bf {#1}} (19{#2}) #3}
\def\ibid#1#2#3{{\it ibid.}~{\bf {#1}} (19{#2}) #3}

\def\szet{{${\scriptstyle \b}$}}
\def\ula{{\un a}}
\def\ulb{{\un b}}
\def\ulc{{\un c}}
\def\uld{{\un d}}
\def\ulA{{\un A}}
\def\ulM{{\underline M}}
\def\cdm{{\Sc D}_{--}}
\def\cdp{{\Sc D}_{++}}
\def\vTheta{\check\Theta}
\def\Pisl{{\Pi\!\!\!\! /}}

\def\fracmm#1#2{{{#1}\over{#2}}}
\def\gg{{\hbox{\sc g}}}
\def\half{{\fracm12}}
\def\ha{\half}

\def\frac#1#2{{\textstyle{#1\over\vphantom2\smash{\raise -.20ex
        \hbox{$\scriptstyle{#2}$}}}}}
\def\fracm#1#2{\hbox{\large{${\frac{{#1}}{{#2}}}$}}}

\def\Dot#1{\buildrel{_{_{\hskip 0.01in}\bullet}}\over{#1}}
\def\dt#1{\Dot{#1}}
\def\uln{{\underline n}}
\def\Tilde#1{{\widetilde{#1}}\hskip 0.015in}
\def\Hat#1{\widehat{#1}}

\def\Dot#1{\buildrel{_{_{\hskip 0.01in}\bullet}}\over{#1}}
\def\dt#1{\Dot{#1}}

\oddsidemargin=0.03in
\evensidemargin=0.01in
\hsize=6.5in
\textwidth=6.5in

\noindent 1.~~{\it Introduction.~~~}Recently there has been considerable
development [1] in understanding $~N=2$~ superstring, especially the
four-dimensional $~(D=4)$\footnotew{We also use $~D=(2,2)$~ to specify
the space-time signature, if necessary.} self-dual supersymmetric
Yang-Mills (SDSYM) and self-dual supergravity
(SDSG) field backgrounds.  The importance of the SDSYM fields has also
attracted attentions in mathematics, due to the strong conjecture [2] that all
integrable models in lower-dimensions are obtained from some dimensional
reductions (DR) of $~D=4$~ self-dual Yang-Mills (SDYM) theory.

        In our recent papers [3-6] we have further generalized this
conjecture from {\it purely bosonic} cases to the
{\it supersymmetric}\footnotew{We use the word ``supersymmetric'' for the
{\it space-time} supersymmetries as ``square-roots'' of translations,
but {\it not} for supergroups or other
fermionic extensions.}~~integrable systems, and established the formulation
of SDSYM theories in $~D=(2,2)$~ space-time, which has more significance
for the integrable systems than the Euclidean signature $~D=(4,0)$.  We
have shown in Ref.~[7] that
some {\it supersymmetric} integrable models, such as super KdV (SKdV)
[8] and super Toda [9] or Liouville theories are actually generated by our
$~N=2$~ SDSYM theory, after appropriate dimensional reductions, as the
first supporting evidence for our conjecture.

        Motivated by these developments, it is a natural next step to
seek other {\it supersymmetric} integrable models in lower-dimensions.
One interesting case is $~N=1$~ supersymmetric Kadomtsev-Petviashvili
(SKP) equations [10].  In this Letter we show that our $~N=1~$ SDSYM
can actually embed this system upon a dimensional reduction with
truncations into $~D=3$, which yields supersymmetric Chern-Simons (SCS)
theory [11].
Based on this result, we will also give some conjectural $~N=2$~ SKP
system.  Additionally sets of exact solutions for the $~N=1$~ SKP as
well as the (conjectural) $~N=2$~ SKP systems are given.

\bigskip\bigskip

\noindent 2.~~{\it DR into $~D=3$~ SCS Theory.~~~}We first establish an
appropriate DR with truncation that yields SCS theories in $~D=3$~ out of the
$~D=(2,2)$~ SDSYM.  Recall that the Bianchi identities (BIds) to
be satisfied for the $~D=(2,2)$~ SDSYM are [3-6]\footnotew{We are using
essentially the same notations as in Ref.~[6] {\it except for}
the {\it hats} to be explained below.}
$$\hat \nabla _{\[ \hat A} \hat F_{\hat B \hat C)}{}^I
- \hat T\du {\[\hat A\hat B|} {\hat D} \hat F\du{\hat D|\hat C)} I \equiv 0 ~~.
\eqno(2.1) $$
with
$$ \hat F\du{\hat\ula\hat\ulb} I = \half \hat \e \du{\hat\ula\hat\ulb}
{\hat\ulc\hat\uld}
\hat F\du{\hat c\hat d} I ~~,
\eqno(2.2) $$
for the self-duality of the superfield strength.  We are using {\it hatted}
indices and fields in $~D=(2,2)$~ superspace, in order to distinguish them from
the $~D=3$~ after the DR.  The indices $~{\scst I,~J,~\cdots}$~ are for
the adjoint representations for the YM gauge group.
In our signature convention $~(+,+,-,-)$, eq.~(2.2) implies
$$\hat F\du {1 2} I = \hat F\du{3 4} I ~~, ~~~~\hat F\du {2 3} I = \hat
F\du{4 1} I ~~, ~~~~ \hat F\du {3 1} I = \hat F\du{4 2} I ~~.
\eqno(2.3) $$
We can impose also the usual simple DR rule
$$ \partial_2 = 0 ~~,
\eqno(2.4) $$
if we regard $~(x^1,x^3,x^4)$~ as the resultant $~D=3$~ coordinates into
which the DR is performed.
Eq.~(2.3) is a necessary condition of the original SD (2.2).  In the process
of the DR, we can further truncate some of the fields as superspace
constraints consistently
with the BIds.  For example, we can put the extra constraints
$$\hat F_{41} {}^I = \hat F_{42}{}^I = \hat F\du{43} I = 0 ~~.
\eqno(2.5) $$
Now (2.4) and (2.5) with (2.3) imply that
$$ F\du{a b} I = 0 ~~,~~~~ {\scst (a,~b~=~1,~3,~4)}~~,
\eqno(2.6) $$
which means the {\it pure-gauge} condition for the YM field
strength in $~D=3$~ after the DR.  Interestingly enough, this system
is nothing else than the
SCS theory [11], and this feature is independent of the number $~N$~ of
supersymmetries.  The consistency with supersymmetry is now clear,
because we know that (2.6) is consistent with the BIds in $~
D=3,\, N\ge 1$~ supersymmetry.

Eventually this DR-truncation yielding the vanishing superfield strength
is similar to what we call ``second type DR'' in our previous paper [7] for
embedding SKdV systems.  The slight difference of the present scheme is
that we have imposed extra on-shell truncation rules (2.5) as superspace
constraints from outside.
This also implies that the $~D=3$~ system (2.3) first obtained from
the $~D=4$~ SDSYM is more general than the SCS theory.

\bigskip\bigskip

\noindent 3.~~{\it Embedding $~N=1$~ SKP Equations.~~~}We now try to
embed the $~N=1$~ SKP system into one of the SCS theories in $~D=3$.
The $~N=1$~ SKP equation is given by [10]
$$\fracm 34 \dery^2 \Psi = - \derx \left[ \dert \Psi +
\fracm 14 \derx^3 \Psi + \fracm 32 \derx (\Psi D\Psi)  \right]~~,
\eqno(3.1) $$
in terms of the $~N=1$~ {\it fermionic} superfield
$$\Psi(t,x,y,\theta) \equiv \phi(t,x,y) + \q u(t,x,y) ~~.
\eqno(3.2) $$
Our $~D=3$~ coordinates are now $~(t,x,y)$.
We are using $~\dert \equiv \partial/ \partial t,~
\derx \equiv \partial/ \partial x,~\dery \equiv \partial/ \partial y$,
and $~D\equiv \partial/\partial\q + \q \derx,~ D^2\equiv\derx$.
The $~\phi\-$field is {\it fermionic}, while $~u$~ is {\it bosonic}.
The corresponding component field equations are
$$\li{&\fracm34 \dery^2 u + \derx \left[ \dert u + \fracm14 \derx^3 u +
3u \derx u + \fracm 32 ( \derx^2 \phi) \phi \right]= 0 ~~,
&(3.3) \cr
& \fracm34 \dery^3\phi + \derx \left[ \dert \phi + \fracm14 \derx^3 \phi
+ \fracm 32 \derx(u \phi) \right] = 0 ~~.
&(3.4) \cr } $$

We can show that eq.~(3.1) can be embedded into a $~D=3,\, N=1$~ SCS theory,
coming from the $~D=4~$ SDSYM.
The $~D=3,~N=1$~ superfield equations of the SCS theory are
$$F_{A B} =0~~, ~~~~{\scst (A,~B,~\cdots~=~t,~x,~y,~\theta)}
\eqno(3.5) $$
with the $~U(1)$~ superfield strength $~F_{A B}$.  Our ans{\" a}tze for
the potential superfields are
$$\li{& A_x = \fracm 34 \derx\dery D\Psi ~~, ~~~~ A_t = \fracm34 \dert
\dery D \Psi ~~,  \cr
& A_y = - \derx D\left[ \dert \Psi + \fracm 14 \derx^3 \Psi + \fracm 32
\derx (\Psi D\Psi)  \right] ~~,
& (3.6) \cr
& \G_\theta = \fracm 34 \derx\dery \Psi ~~.
& (3.7) \cr } $$

        Eq.~(3.5) can be easily confirmed for these ans{\" a}tze.
First, $~F_{t x} = 0$~ is clear, because $~A_x~$ and ~$A_t~$ have respectively
$~\derx$~ and $~\dert$~ acting on the same function ~$ (3/4) D\dery
\Psi$.  Second, $~F_{x y}$~ and $~F_{t y}$~ vanishes by the use of
superfield equation (3.1).  For example,
$$F_{x y} = - \derx D\left[ \, \fracm 34 \dery^2 \Psi
+ \derx\left\{ \dert
\Psi + \fracm 14 \derx^3 \Psi + \fracm 32 \derx (\Psi D \Psi)
\right\} \right] = 0 ~~,
\eqno(3.8) $$
and similarly for $~F_{t y}=0$.
The remaining components are $~F_{t \q},~F_{x\q}$~ and $~F_{y\q}$, where
the first two vanish identically, while
$$F_{y \q} = D\left[\fracm 34 D\dery^2 \Psi + D \derx \left\{
\dert \Psi + \fracm 14 \derx^3 \Psi + \fracm 32 \derx (\Psi D\Psi)
\right\}  \right]  = 0 ~~
\eqno(3.9) $$
vanishes by the use of (3.1).

        If we review our ans{\" a}tze (3.6) and (3.7), we see that
$$ A_y = \fracm 34 D\dery^2 \Psi ~~
\eqno(3.10) $$
holds {\it on-shell}, using (3.1).  In other words, we have the universal
{\it on-shell} relation
$$ \G_A \equiv (A_a,\G_\q) = D_A \left[ \fracm 34 \dery D \Psi
\right]~~,~~~~{\scst (a,~b,~\cdots ~=~ t,~x,~y)} ~~.
\eqno(3.11)$$
It is now much more transparent why the superfield strength $~F_{A B}$~
vanishes, because all the potential superfields are total
super-gradients.
Notice, however, that we need the expression (3.6) in order to supply
the information about
the SKP equation, otherwise all the components vanish {\it identically}
without yielding any equation.

\bigskip\bigskip

\noindent 4.~~{\it Conjectural $~N=2$~ SKP Equations.}~~~Once we
have understood how the $~N=1$~ SKP equations can be embedded into $~N=1$~ SCS
system [11], it is easier to extrapolate the result to conjecture
possible $~N=2$~ SKP equations.  Our strategy is based on the two
main principles:

\noindent (i)  The $~N=2$~ SKP superfield equation contains the
derivative $~\dery$~ only in a second derivative term $~\approx\dery^2$~
acting on an $~N=2$~ superfield $~\Phi$, and if $~\dery^2\-$term is set
to zero, the superfield equation is reduced to the usual $~N=2$~ SKdV
superfield equation [8].

\noindent (ii)  The r.h.s.~of (3.1) for the $~N=1$~ SKP is to be replaced by
the $~N=2$~ SKdV equation [8] with an overall $~\derx\-$derivative in front
providing the right mass-dimension.

\noindent These principles are from the empirical fact that $~N=1$~
SKP equations [10] contain $~N=1$~ SKdV equations, as their sub-cases.

Under these principles, there is a unique (up to inessential factors)
equation we can put:
$$\eqalign{3 \dery^2 \Phi = \, - \derx \bigg[\, & \dert \Phi + \derx^3
\Phi - 3 \derx ( \Phi D_1 D_2 \Phi)  \cr
& - \half (\a-1) \derx (D_1 D_2 \Phi^2) - 3\a \Phi^2 \derx \Phi \, \bigg]
{}~~, \cr}
\eqno(4.1) $$
where $~\Phi$~ is a {\it bosonic} $~N=2$~ scalar superfield
mimicking the $~N=2$~ SKdV system [8]:
$$\Phi(t,x,y,\theta^1,\theta^2)  = w(t,x,y) + \theta^1 \xi_1(t,x,y) + \theta^2
\xi_2 (t,x,y) + \theta^2 \theta^1 u(t,x,y) ~~,
\eqno(4.2) $$
except that this $~\Phi\-$superfield can now depend also on $~y$.
The requirement of the
{\it integrability} of $~N=2$~ SKdV as its sub-theory, the constant $~\a~$ is
to be $~\a=-1,~2$~ or $~4$~ [8].  The $~D_i$~ are the usual
$~N=2$~ supercovariant derivatives:
$$ D_i \equiv \fracmm{\partial}{\partial\q^i} + \q^i
\fracmm{\partial}{\partial x} ~~, ~~~~ D_1 {}^2 = D_2{}^2 =
\fracmm{\partial}{\partial x} ~~, ~~~~ \{ D_1,\, D_2\} = 0~~.
\eqno(4.3) $$

        Our $~N=2$~ SCS superspace condition is that
the $~U(1)$~ superfield strength is to vanish:
$$F_{A B} = 0 ~~, ~~~~{\scst (A,~B,~\cdots~=~ t,~x,~y,~\q^1,~\q^2)}~~.
\eqno(4.4) $$
Now our ans{\" a}tze for the superpotentials are
$$ \eqalign{& A_x = 3 \derx\dery \Phi~~,  ~~~~ A_t = 3
\dert\dery \Phi ~~, \cr
& A_y = - \derx \bigg[\, \dert \Phi + \derx^3 \Phi - 3 \derx ( \Phi D_1
D_2 \Phi ) \cr
& ~~~~~ ~~~~~ ~~~~~ - \half (\a-1) \derx (D_1 D_2 \Phi ^2) - 3\a \Phi^2 \derx
\Phi \, \bigg] ~~, \cr
& \G _i = 3 \dery D_i \Phi ~~, \cr }
\eqno(4.5) $$
where $~\G_i$~ stands for $~\G_{\q^i}$.

        We can easily confirm that the satisfaction of (4.4) for each
indices of $~{\scst A B}$.  In fact, $~F_{x t}~$,~$F_{x \q^i}, ~F_{t
\q^i}$~ and $~F_{\q^i \q^j}$~ all vanish {\it identically}, while $~F_{y
t}, ~F_{y x}$~ and $~F_{y \q^i}$~ are zero, {\it only when} the field equation
(4.1) is used.

        The feature we had in (3.11) for the embedding of $~N=1$~
SKP is re-encountered in this system.  As a matter of fact, we can see that the
relations
$$ \G_A \equiv (A_a,\G_i) = D_A \left( 3 \dery \Phi \right) ~~,
{}~~~~{\scst (a,~b,~\cdots ~=~ t,~x,~y)}
\eqno(4.6) $$
hold {\it on-shell}, which make the above confirmation much easier and
explicit.

        Even though we have not performed its confirmation,
we have enough reason to conjecture the {\it integrability} of our
$~N=2$~ SKP equation (4.1).

\bigskip\bigskip

\noindent 5.~~{\it Exact Solutions for $~N=1$~ and Conjectural $~N=2$~
SKP Equations.}~~~As a next
interesting task, we give a set of exact solutions for the $~N=1$~
SKP equations (3.3) and (3.4).  First, the resemblance of the two component
equations (3.3) and (3.4) is suggestive that the two fields $~u$~
and $~\phi$~ share the same function as exact solutions.  Motivated by
this observation, we put the ansatz for $~\phi$:
$$\phi = \zeta u~~, ~~~~ (\zeta ^2 \equiv 0)~~,
\eqno(5.1) $$
where $~\zeta$~ is a constant Grassmann spinor.  By
simple algebra, we easily see that eq.~(3.4) is reduced into exactly the
same differential equation as (3.3), {\it without} the $~\phi^2\-$term.
In other words, the only new differential equation to be solved is the purely
{\it bosonic} one:
$$ 3 \dery^2 u + \derx \left[4 \dert u + \derx^3 u +  12 u \derx u
\right] = 0 ~~.
\eqno(5.2) $$
A wave-like solution to this equation is easily obtained, as usual [12,13],
by assuming the special dependence of $~u$~ on the variables:
$$ u (x,y,t) = f(ax+by+c t) \equiv f(\xi) ~~, ~~~~ \xi \equiv ax+by+c t~~.
\eqno(5.3) $$
Accordingly this $~f(\xi)$~ satisfies
$$\left[ a^4 f'' +6 a^2 f^2 +(4ac+3b^2) f \right] '' = 0~~ ,
\eqno(5.4) $$
where each prime denotes a derivative $~d/d\xi$.
This equation can be integrated under the usual boundary condition
$~f\rightarrow 0,~\hbox{as}~|\xi| \rightarrow \infty$:
$$ \li{f & =  -\fracmm{4ac+ 3b^2}{4a^2\cosh^2 \left[
\fracmm{\sqrt{|4ac+3b^2|}}{2a^2}(\xi-\xi_0)\right]}
&(5.5a) \cr
& = 2 \derx^2 \ln \left[\, \cosh \left\{ \fracmm{\sqrt{|4ac + 3b^2|}}{2a^2}
(\xi - \xi_0) \right\} \right] ~~,
&(5.5b) \cr} $$
where $~\xi_0\equiv a x_0+b y_0 + c t_0$~ is an arbitrary constant.
We also assumed $~4 a c + b^2 < 0$, because this case includes
the solitary wave solution in its $~N=1$~ SKdV sub-theory with
$~a=1,~b=0,~c= - |c| < 0$~ [12].

        Once a bosonic solution for $~u$~ has been obtained, the fermionic
solution is straightforward {\it via} (5.1):
$$\li{ &u = - \fracmm{4ac+ 3b^2}{4a^2\cosh^2 \left[
\fracmm{\sqrt{|4ac+3b^2|}}{2a^2}\{a(x-x_0) + b(y-y_0) +
c(t-t_0)\}\right]}~~,
&(5.6) \cr
& \phi = - \fracmm{4ac+ 3b^2}{4a^2\cosh^2 \left[
\fracmm{\sqrt{|4ac+3b^2|}}{2a^2}\{a(x-x_0) + b(y-y_0) +
c(t-t_0)\}\right]} \zeta  ~~.
&(5.7)  \cr } $$

        Needless to say, our $~N=1$~ SKP system has the $~N=1$~ SKdV
system [8] as its sub-case, obtained by requiring the
independence of all the fields on $~y$~ [10].  Actually this provides a
good double-check, and in fact we see the SKdV solutions [12] are
re-obtained after truncating the $~y\-$dependence of the solutions
(5.6) and (5.7).  Additionally, if we require the independence of $~t$~ in
the solution to (5.2), the system is reduced to what is called
Boussinesq equation.  Accordingly, our system above is regarded as
$~N=1$~ supersymmetric Boussinesq equations, and its solutions are (5.6)
and (5.7) with $~c=0$.

It is worthwhile to mention the existence of another
class of solutions in terms of $~g\-$dimensional
elliptic $~\vartheta\-$functions on an arbitrary Riemann surface $~\G$~
[13], if we release our boundary
condition $u \rightarrow 0~(|x| \rightarrow\infty)\,$:
$$u = 2 \derx^2 \ln \Big[ \vartheta \Big( \Vec U x + \Vec V y + \Vec W
t -\Vec I \Big| B \Big) \Big] + C ~~,
\eqno(5.8) $$
where
$$\eqalign{&\vartheta (\Vec z\big| B) \equiv \vartheta \pmatrix{\Vec 0 \cr \Vec
0 \cr} (\Vec z\big| B) ~~, \cr
&\vartheta \pmatrix{\Vec\a \cr \Vec\b \cr }(\Vec z\big| B)
\equiv \sum_{\Vec n\in{{}_{\bf Z}}^g}
\exp \Big[\pi i (\Vec n+ \Vec \a)\cdot B (\Vec n+ \Vec\b)
+ 2\pi i (\Vec n+\Vec\a)\cdot(\Vec z + \Vec\b) \Big] ~~. \cr }
\eqno(5.9) $$
Here $~g$~ is the genus of $~\G$, while
$~\Vec U, ~\Vec V, ~ \Vec W$~ are the $~g\-$dimensional vectors
of $~b\-$periods of
some normalized Abelian integrals of the second kind with the poles at
the marked point $~P_0\in \G$, and $~B$~ is a the $~g\times g$~ matrix of
$~b$~ periods of the surface $~\G$~ in some canonical bases [13].  Since
this solution is based on the elliptic
$~\vartheta\-$function which is {\it periodic} in its variable,
the solutions does {\it not} satisfy the usual boundary condition $~u
\rightarrow 0 ~(|x| \rightarrow \infty)$.  However, such a solution
may have special significance related to partition functions at higher-loop
in string theory, because it is written for general genus $~g$.
The similarity between (5.5b) and (5.8) is also
suggestive that other solutions to (5.2) share the common structure:
$~u = 2 \derx^2 \ln F$~ for some function $~F$.

        The above solution has an interesting form in superfields:
$$ \Psi = \left (\zeta + \theta \right) u = \Omega u ~~, ~~~~
\Omega \equiv  \zeta + \theta ~~.
\eqno(5.10) $$
It is interesting that not only the solution (5.5) but also any other
solution for $~u$~ to (5.2) can be utilized as the common-factor function
in (5.7).  This sort of exact solutions can be also obtained by
supertranslations of the purely bosonic solution of $~u$, as is easily
understood by applying supertransformation to the latter.
More general exact solutions for the SKP are rather difficult to get,
but we refer readers to Ref.~[14] for exploring other solutions.

        We can follow the same strategy for getting exact solutions for
our optional $~N=2$~ SKP equation (4.1).  We put the ansatz
$$\Phi = \Xi u ~~, ~~~~ \Xi = \zeta_2 \zeta_1 + \theta^i \zeta_i +
\theta_2 \theta_1 ~~, ~~~~(\zeta_1 ^2 = \zeta_2 ^2 = \{ \zeta_1,\zeta_2 \} =
0)~~.
\eqno(5.11) $$
It is now straightforward to confirm that $~u$~ satisfies
$$ 3 \dery^2 u + \derx\left[ \derx^3 u - 6 u \derx u - \dert u  \right]
= 0 ~~,
\eqno(5.12) $$
which is again solved by the function like (5.6):
$$ u = \fracmm{- ac+ 3b^2}{2a^2\cosh^2 \left[
\fracmm{\sqrt{|ac-3b^2|}}{2a^2}\{a(x-x_0) + b(y-y_0) +
c(t-t_0)\}\right]} ~~.
\eqno(5.13) $$
Eventually the exact solution is (5.11) with $~u$~ in (5.13).

\bigskip\bigskip

\noindent 6.~{\it Concluding Remarks.~~~}In this Letter we have shown how
the $~N=1$~ SKP equations [10] are embedded into the $~D=(2,2),\,N=1$~ SDSYM
[3-6] {\it via} $~D=3,\,N=1$~ SCS theory.
We have also given some conjectural $~N=2$~ SKP equations, which are
likely to be integrable.  We have further given some exact solutions of
these equations, especially with the fermionic solutions, sharing
the same functions as the {\it purely} bosonic solutions, as is understood by
supersymmetric covariance.  Similar feature is also found for the
$~N=2$~ conjectural integrable SKP equations.  We have also mentioned
the interesting solutions in terms of elliptic $~\vartheta\-$functions,
which may be physically related to partition functions in string
theory.

        Another consequence of our results is that the
{}~$D=3$~ SCS theory [11] played an important role for embedding these SKP
equations.\footnotew{It seems to us after lots of trials that there is
{\it no} other DR that can {\it bypass} the SCS theories, at least
in the {\it on-shell} formulation.}~~In other words, we
have shown an important link between the SKP equations and
{\it supersymmetric} purely topological theories.  Since SKdV equations are
contained as a sub-theory in the SKP systems, our results
further imply the close relationship between the SKdV equations and
SCS theories.
As a matter of fact, in our previous paper [7] we have suggested a
possible connection among the SKdV equations and SCS equations.  In
this Letter we gave the explicit supporting evidence of these
remarkable connections {\it via} SKP systems, containing the SKdV
equations.

        Before concluding, we mention the following point about
{\it extended} SKdV systems embedded in the $~D=(2,2)$~ SDSYM.
There has been recently some progress in the direction of
extending supersymmetries to $~N=4$~ for SKdV equations [15],
based on what is called $~N=4$~ harmonic superspace.
In this system, we need at least six
scalars to form a multiplet, and to accommodate this system in
$~D=(2,2)$~ SDSYM, we have to start with $~N=4$~ maximal irreducible SDSYM
system related to the $~N=2$~ superstring first presented by Siegel [16].
To be more specific, recall first the field content of the $~N=4$~ SDSYM
[6,16] $~(A\du\ula I, G\du{\ula\ulb} I, \r^I, \Tilde\l^I, S\du{\hat i} I,
T\du{\hat i} I)$~ in the notation of Ref.~[6].
If we compare this fields with the $~N=4$~ SKdV, we
realize that the multiplier fields $~G\du{\ula\ulb}I$~ and $~\r^I$~
are to be truncated to embed the latter into the former.
However, as is easily revealed by simple inspection on the closure of
supersymmetry,
we can {\it not} delete these multiplier fields, maintaining the $~N=4$~
supersymmetry.  Thus we have to truncate some of the supersymmetries down to
$~N\le 2$, which {\it lack} enough scalars to embed the $~N=4$~ SKdV
[15].  In other words, it is {\it unlikely} that the $~N=4$~ SKdV system
proposed in Ref.~[15] has any link with the $~N=2$~ superstring theory
or its consistent background of the $~N=4$~ SDSYM [16] .

        There are lots of possible future developments based on our
results.  For example, we can investigate the {\it topological} significance
of the link between the SKP equations and SCS theories [11].  We can also
seek other {\it integrable} systems with {\it extended} supersymmetries
into $~D=4$~ SDSYM theories.
The quantum aspects of these integrable theories can be understood
better, by investigating the underlying $~N=2$~ superstring [1,16]
of the SDSYM fields.  As a matter of fact, in our recent paper [17] we
have shown an important link between the $~N=2$~ superstring and the
two-dimensional topological field theory, especially a direct
relationship of $~W_\infty\-$gravity to the $~N=2$~ superstring is
elucidated.
The possibility of other integrable models
embedded in SDSG may be equally attractive related to
$~W_\infty\-$(super)gravity.

\bigskip\bigskip

We are grateful to S.J.~Gates, Jr.~for providing interesting ideas and
important suggestions.  We are also indebted to D.~Depireux for information
about exact solutions.

\vfill\eject

\refs
\small

\items{1} H. Ooguri and C. Vafa, \mpl{5}{90}{1389};
\np{361}{91}{469}; \ibid{367}{91}{83};
H.~Nishino and S.J.~Gates, Jr., \mpl{7}{92}{2543}.

\items{2} M. F. Atiyah, unpublished;
R. S. Ward, Phil.~Trans.~Roy.~Lond.~{\bf A315} (1985) 451;
N. J. Hitchin, Proc.~Lond.~Math.~Soc.~{\bf 55} (1987) 59.

\items{3} S.V.~Ketov, S.J.~Gates and H.~Nishino, Maryland preprint,
UMDEPP 92--163, submitted to Phys.~Lett.~B.

\items{4} H. Nishino, S. J. Gates, Jr. and S. V. Ketov,
Maryland preprint, UMDEPP 92--171, submitted to Phys.~Lett.~B.

\items{5} S. J. Gates, Jr., H. Nishino and S. V. Ketov, \pl{297}{92}{99}.

\items{6} S.J.~Ketov, H.~Nishino and S.J.~Gates, Jr., Maryland preprint,
UMDEPP 92--211, to appear in Nucl.~Phys.~B.

\items{7} S.J.~Gates, Jr.~and H.~Nishino, \pl{299}{93}{255}.

\items{8} P.~Mathieu, Jour.~Math.
Phys.~{\bf 29} (1988) 2499;
C.A.~Laberge and P.~Mathieu, Phys.~Lett.~
\item{  } {\bf 215B} (1988) 718;
T.~Inami and H.~Kanno, \ijmp{A7}{92}{419}.

\items{9} E.~Ivanov and S.~Krivonos, Lett.~Math.~Phys. {\bf 7} (1983) 523;
L.A.~Leites, M.V.~Saveliev and V.V.~Serpukgov, ``Embeddings of
of $~OSp(N/2)$~ and Associated Non-linear Supersymmetric Equations'',
{\it in} Proc.~Third Yurmala Seminar (USSR 22-24 May 1985);
M.A.~Olshanetsky, \cmp{88}{83}{63};
J.~Evans and T.~Hollowood, \np{352}{91}{723}.

\items{10} Y.~Manin and A.O.~Radul, \cmp{98}{85}{65}.

\items{11} W.~Siegel, \np{156}{79}{135};
S.J.~Gates, M.T.~Grisaru, M.~Ro{\v c}ek and W.~Siegel, ``{\it
Superspace}'', (Benjamin/Cummings, Reading MA, 1983), page~27;
N.~Sakai and Y,~Tanii, \ptp{83}{90}{968};
H.~Nishino and S.J.~Gates, Jr., Maryland preprint, UMDEPP 92-060, to
appear in Int.~Jour.~Mod.~Phys.

\items{12} See {\it e.g.}, P.G.~Drazin and R.S.~Johnson, {\it ``Solitons: An
Introduction''}, Cambridge Univ.~Press (1989).

\items{13} I.M.~Krichever, Functional Analysis and Its Applications {\bf 11}
(1977) 12; W.B.~Mateev and A.O.~Smirnov, Lett.~Math.~Phys.~{\bf 14}
(1987) 25.

\items{14} M.~Mulase, Inventiones Mathematicae {\bf 92} (1988) 1.

\items{15} F.~Deluc and E.~Ivanov, ENS-Lyon preprint, ENSLAPP-L-415-92,
(Dec.~1992).

\items{16} W.~Siegel, Stony Brook preprint, ITP-SB-92-24 (May, 1992).

\items{17} H.~Nishino, Maryland preprint, UMDEPP 93-144 (Feb.~1993).

\end{document}